\documentclass[letterpaper, preprint, paper,11pt]{AAS}	

\usepackage{bm}
\usepackage{amsmath, amsfonts}
\usepackage{subfigure}
\usepackage[colorlinks=true, pdfstartview=FitV, linkcolor=black, citecolor= black, urlcolor= black]{hyperref}
\usepackage{overcite}
\usepackage{footnpag}			      	

\newcommand{\destiny}{DESTINY$^+$}

\PaperNumber{23-222}

\begin{document}

\title{Low-Thrust Many-Revolution Trajectory Design Under Operational Uncertainties for \destiny \ Mission}

\author{Naoya Ozaki\thanks{Ph.D., Associate Professor, Department of Spacecraft Engineering, Institute of Space and Astronautical Science, Japan Aerospace Exploration Agency, Kanagawa 252-5210, Japan; ozaki.naoya@jaxa.jp},
Yuki Akiyama\thanks{Ph.D., Researcher, Space Tracking and Communications Center, Japan Aerospace Exploration Agency, Ibaraki 305-8505, Japan; akiyama.yuuki@jaxa.jp},
Akira Hatakeyama\thanks{Master's student, Department of Applied Mechanics and Aerospace Engineering, Waseda University, Waseda University, 1-104 Totsukamachi, Shinjuku-ku, Tokyo, 169-8050, Japan.},
Shota Ito\thanks{Master's student, Department of Aerospace Engineering, Tokyo Metropolitan University, Tokyo Metropolitan University, 6-6, Asahigaoka, Hino, Tokyo, 191-0065, Japan.},
Takuya Chikazawa\thanks{Ph.D. student, Department of Advanced Energy, The University of Tokyo, The University of Tokyo, 7-3-1, Hongo, Bunkyo-ku, Tokyo, 113-8656, Japan.},
\ and Takayuki Yamamoto\thanks{Ph.D., Senior Researcher, DESTINY$^+$ Project Team, Institute of Space and Astronautical Science, Japan Aerospace Exploration Agency, Kanagawa 252-5210, Japan; yamamoto.takayuki@jaxa.jp}
}

\maketitle{} 		

\begin{abstract}
\destiny is a planned JAXA medium-class Epsilon mission from Earth to deep space using a low-thrust, many-revolution orbit. Such a trajectory design is a challenging problem not only for trajectory design but also for flight operations, and in particular, it is essential to evaluate the impact of operational uncertainties to ensure mission success. In this study, we design the low-thrust trajectory from Earth orbit to a lunar transfer orbit by differential dynamic programming using the Sundman transformation. The results of Monte Carlo simulations with operational uncertainties confirm that the spacecraft can be successfully guided to the lunar transfer orbit by using the feedback control law of differential dynamic programming in the angular domain.
\end{abstract}

\section{Introduction}

\begin{figure}[h!]
\centering
\includegraphics[width=0.7\textwidth]{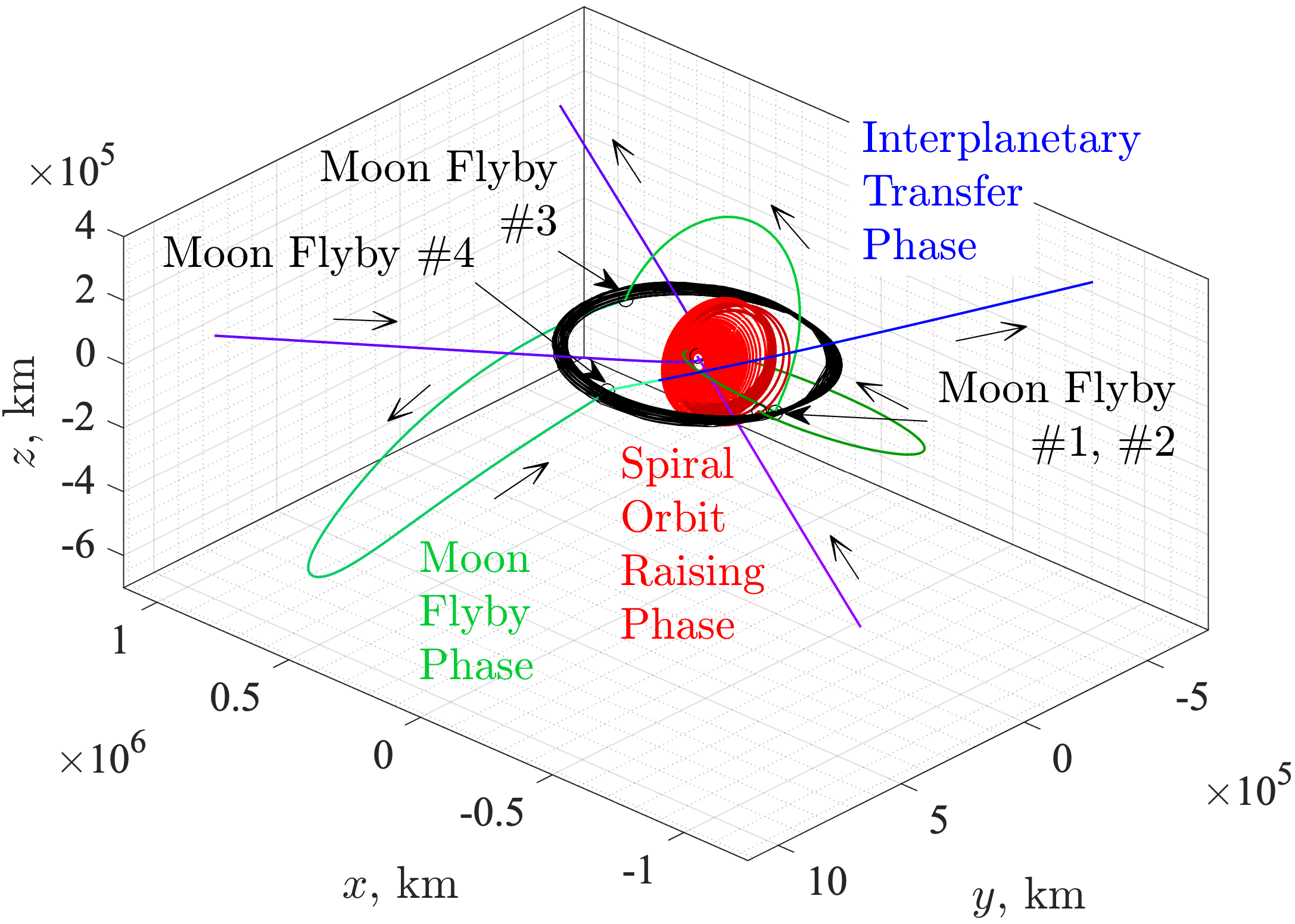}
\caption{Baseline near-Earth trajectory of \destiny\ in the Earth-centered ECLIPJ2000 inertial frame.}
\label{fig:baseline_earthescape_trajectory}
\end{figure}

%
%
%
Deep space exploration is undergoing a transformative phase thanks to low-cost, frequent missions. In this context, JAXA is leading the development of the \destiny (Demonstration and Experiment of Space Technology and INterplanetary voYage, Phaethon fLyby and dUst Science) mission.\cite{Ozaki2022} Expected to launch in the mid-2020s, \destiny will ride on the low-cost Epsilon S launch vehicle. The Epsilon S is equipped to place the spacecraft into a near-geostationary transfer orbit. The spacecraft will then use solar electric propulsion to gradually raise its orbit and follow a spiral trajectory into deep space, as shown in the baseline near-Earth trajectory in Fig. \ref{fig:baseline_earthescape_trajectory}. This groundbreaking endeavor will be the world's first mission to transition from near-geostationary transfer orbit to deep space using a low-thrust propulsion system. Achieving this milestone depends on the use of an efficient solar electric propulsion system powered by an upgraded variant of the $\mu 10$ ion thruster on the Hayabusa2 spacecraft, complemented by lightweight, high-efficiency solar panels. 

%
%
%
%
%
%
%
Due to the nature of low-thrust propulsion systems, this mission requires a many-revolution trajectory design to leave Earth orbit into deep space. One of the state-of-the-art techniques for designing low-thrust many-revolution trajectories is based on Hybrid Differential Dynamic Programming (HDDP)\cite{Lantoine2012p1, Lantoine2012p2}. HDDP is a variant of Differential Dynamic Programming (DDP)\cite{Jacobson1970, Lin1991p1, Lin1991p2, Whiffen2002} and computes the local optimal control by solving a second order expansion of the Bellman equation. Aziz et al.\cite{Aziz2017} have applied a Sundman transformation to shift the independent variable from the time domain to an orbital angle domain, resulting in a significant improvement in the effectiveness of the HDDP approach for low-thrust many-revolution trajectory design. In addition to the trajectory optimization challenges, the flight mission operation must consider the uncertainties resulting from missed thrust, navigation errors, and low-thrust execution errors. Although researchers have studied the stochastic trajectory design approach in recent years\cite{Ozaki2018, Greco2022, Ozaki2020a, Oguri2022}, there is no established method for optimizing low-thrust many-revolution trajectories to account for operational uncertainties. At the least, the assessment of the impact of uncertainties is an essential study for the guarantee of the success or failure of the mission.

%
%
%
In this paper, we first design a low-thrust, many-revolution trajectory to the Moon using the HDDP techniques with the Sundman transformation. To evaluate the effects of the uncertainties, we perform the Monte Carlo simulation employing four different control policies to correct the perturbed trajectory. Among the four policies, the local linear feedback control policy provided by HDDP under the Sundman transformation records the best performance to guide the spacecraft towards the moon. Finally, we discuss the usability of the proposed method and strategies to improve robustness to uncertainties through numerical results.

\section{Hybrid Differential Dynamic Programming with Sundman Transformation}

This section summarizes the basics of the HDDP algorithm and the Sundman transformation, and introduces the notation and assumptions necessary for the following analysis. Our HDDP implementation is based on the work of Lantoine and Russell\cite{Lantoine2012p1, Lantoine2012p2}, Aziz et al.\cite{Aziz2017}


%
%
\subsection{Problem Formulation}

We consider the single-phase, constrained, discrete-time optimal control problem as follows. For a state vector $\bm{x}_k\in\mathbb{R}^{n_x}$ and control vector $\bm{u}_k\in\mathbb{U}\subseteq\mathbb{R}^{n_u}$, the objective function is defined as
\begin{equation}
	J(\bm{x}_0, \{\bm{u}_k\}_{k\in\mathbb{N}_{N}}) := \sum_{k=0}^{N}l_k(\bm{x}_k, \bm{u}_k) + \phi(\bm{x}_{N+1})\label{eq:DDP_objective_function}
\end{equation}
The dynamical system is defined as
\begin{equation}
	\bm{x}_{k+1} = \bm{f}_k(\bm{x}_k, \bm{u}_k), \ \ \ k\in\mathbb{N}_{N}.\label{eq:DDP_dynsys}
\end{equation}
by numerically solving the continuous-time equation of motion of the spacecraft. The optimal control problem is subject to the stage constraints
\begin{equation}
\bm{g}_k(\bm{x}_k, \bm{u}_k) \leq \bm{0}
\end{equation}
and the phase constraints
\begin{equation}
    \bm{\psi}(\bm{x}_{N+1}) = \bm{0}.
\end{equation}

To deal with the phase constraints, we replace the phase cost function $\phi(\bm{x}_{N+1})$ by the augmented Lagrangian cost function\cite{Lin1991p1, Lantoine2012p1}
\begin{equation}
\tilde{\phi}(\bm{x}_{N+1}, \bm{\lambda}) := \phi(\bm{x}_{N+1}) + \bm{\lambda}^T \bm{\psi} (\bm{x}_{N+1}) + \sigma \|\bm{\psi} (\bm{x}_{N+1})\|^2
\end{equation}
where $\bm{\lambda}\in\mathbb{R}^{n_\lambda}$ is a Lagrange multiplier and $\sigma\in(0,\infty)\subset \mathbb{R}$ is the penalty parameter introduced to improve the convergence\cite{Lantoine2012p1}.

\subsection{Discrete-time Bellman Equation}

Dynamic Programming (DP) derives its foundations from Bellman's Principle of Optimality\cite{Bellman1957}, which states that "an optimal policy has the property that whatever the initial state and initial decision (i.e., control) are, the remaining decisions must constitute an optimal policy with regard to the state resulting from the first decision". Consequently, rather than directly optimizing the objective function (\ref{eq:DDP_objective_function}), DP takes the characteristic approach of instantiating a cost-to-go function.
\begin{equation}
V_k(\bm{x}_k, \{\bm{u}_j\}_{j\in\mathbb{N}_{[k:N]}}, \bm{\lambda}) := \sum_{j=k}^{N}l_j(\bm{x}_j, \bm{u}_j) + \tilde{\phi}(\bm{x}_{N+1}, \bm{\lambda}) \label{eq:DDP_cost2go_function}
\end{equation}
and finds the optimal control vectors $\{\bm{u}_j^*\}_{j\in\mathbb{N}_{[k:N]}}$ to minimize the cost-to-go function $V_k(\cdot)$. Note that in Eq.(\ref{eq:DDP_cost2go_function}), $\{\bm{x}_j\}_{j\in\mathbb{N}_{[k+1:N]}}$ are determined from $\bm{x}_k$ and the control vectors  $\{\bm{u}_j\}_{j\in\mathbb{N}_{[k:N]}}$ through Eq.(\ref{eq:DDP_dynsys}). The optimal cost-to-go function is defined as
\begin{equation}
V_k^*(\bm{x}_k, \bm{\lambda}) := \min_{\{\bm{u}_j\}_{j\in\mathbb{N}_{[k:N]}}\in\mathcal{U}_{k:N}} V_k(\bm{x}_k, \{\bm{u}_j\}_{j\in\mathbb{N}_{[k:N]}}, \bm{\lambda})\label{eq:DDP_optimal_cost2go}
\end{equation}
where $\mathcal{U}_{k:N}$ is the set of admissible control vectors.

Substituting Eq.(\ref{eq:DDP_cost2go_function}) to Eq.(\ref{eq:DDP_optimal_cost2go}) derives a recursive optimization problem to find the optimal control vector $\bm{u}_k^*$ as
\begin{equation}
V_{k}^*(\bm{x}_k, \bm{\lambda}) = \min_{\bm{u}_k\in\mathcal{U}_k} \left[ l_k(\bm{x}_k, \bm{u}_k) + V_{k+1}^*(\bm{x}_{k+1}, \bm{\lambda}) \right]\label{eq:DDP_bellman_equation}
\end{equation}
where $\mathcal{U}_k$ is a set of admissible control vectors. This equation is called the \textit{Bellman equation}.

%
%
\subsection{Discrete-time Differential Dynamic Programming}

DDP is a numerical method for iteratively solving DP in the neighborhood of a reference trajectory, denoted as $\{\bar{\bm{x}}_k\}_{k\in\mathbb{N}_{N+1}}$, $\{\bar{\bm{u}}_k\}_{k\in\mathbb{N}_{N}}$, and $\bar{\bm{\lambda}}$. By expanding the Bellman equation (\ref{eq:DDP_bellman_equation}) to a second order approximation around the reference trajectory, and then optimizing the quadratic expansion of the cost-to-go function with respect to $\delta \bm{u}_k$ (where $\delta \bm{u}_k = \bm{u}_k - \bar{\bm{u}}_k$), an optimal control correction yields
\begin{equation}
\delta \bm{u}_k^* = \bm{\alpha}_k + \beta_k \delta \bm{x}_k + \gamma_k \delta \bm{\lambda}\label{eq:DDP_control_correction}
\end{equation}
where this control policy is associated with a linear feedback law with respect to state variation.
The coefficients $\bm{\alpha}_k$, $\beta_k$ and $\gamma_k$ manifest as functions of $V_{k+1}^*(\bm{x}_{k+1}, \bm{\lambda})$ with their first and second order derivatives. The coefficients $\bm{\alpha}_k$, $\beta_k$ and $\gamma_k$ at all $k$ th stages can be determined by a backward sweep calculation. Applying Eq.(\ref{eq:DDP_control_correction}) in the forward propagation leads to the updating of the control vectors and the reference trajectory. DDP performs an iterative process of successive forward and backward sweeps. This iterative scheme continues until trajectory convergence is achieved, typically measured by metrics such as reduction in total cost below a predetermined threshold.

%
%
\subsection{Quadratic Expansion of Bellman Equation}

Let us derive the quadratic expansion of the Bellman equation (\ref{eq:DDP_bellman_equation}) around the reference trajectory $(\bar{\bm{x}}_k, \bar{\bm{u}}_k, \bar{\bm{\lambda}})$. As the first step, let us expand $l_k(\cdot)$ and $\bm{f}_k(\cdot)$ around $(\bar{\bm{x}}_k, \bar{\bm{u}}_k, \bar{\bm{\lambda}})$. We neglect the subscript $k$ and use the Cartesian tensor notation for convenience. The variations of $\bm{x}$, $\bm{u}$, and $\bm{\lambda}$ are therefore defined as
\begin{align}
\delta x_i &:= x_i - \bar{x}_i,\\
\delta u_i &:= u_i - \bar{u}_i,\\
\delta \lambda_i &:= \lambda_i - \bar{\lambda}_i,
\end{align}
and the quadratic expansions of $l(\bm{x}, \bm{u})$ and $\bm{f}(\bm{x}, \bm{u})$ are
\begin{align}
f_i(\bm{x}, \bm{u}) &\simeq f_i(\bar{\bm{x}}, \bar{\bm{u}}) + \frac{\partial f_i}{\partial x_j}\delta x_j + \frac{\partial f_i}{\partial u_j}\delta u_j \nonumber\\
& \ \ \ + \frac{1}{2}\frac{\partial^2 f_i}{\partial x_j \partial x_h}\delta x_j \delta x_h + \frac{\partial^2 f_i}{\partial x_j \partial u_h}\delta x_j \delta u_h + \frac{1}{2}\frac{\partial^2 f_i}{\partial u_j \partial u_h}\delta u_j \delta u_h\label{eq:DDP_quadratic_expansion_of_f}\\
l(\bm{x}, \bm{u}) &\simeq l(\bar{\bm{x}}, \bar{\bm{u}}) + \frac{\partial l}{\partial x_j}\delta x_j + \frac{\partial l}{\partial u_j}\delta u_j \nonumber\\
& \ \ \ + \frac{1}{2}\frac{\partial^2 l}{\partial x_j \partial x_h}\delta x_j \delta x_h + \frac{\partial^2 l}{\partial x_j \partial u_h}\delta x_j \delta u_h + \frac{1}{2}\frac{\partial^2 l}{\partial u_j \partial u_h}\delta u_j \delta u_h\label{eq:DDP_quadratic_expansion_of_l}
\end{align}
where all the derivatives are evaluated at $(\bar{\bm{x}},\bar{\bm{u}})$. Also, we redefine $V_{k+1}^*(\bm{x}_{k+1}, \bm{\lambda})$ as $V^+(\bm{x}^+, \bm{\lambda})$ and expand it quadratically as
\begin{align}
V^+(\bm{x}^+, \bm{\lambda}) &\simeq V^+(\bar{\bm{x}}^+, \bar{\bm{\lambda}}) + \frac{\partial V^+}{\partial x_j^+}\delta x_j^+ + \frac{\partial V^+}{\partial \lambda_j}\delta \lambda_j \nonumber\\
& \ \ \ + \frac{1}{2}\frac{\partial^2 V^+}{\partial x_j^+ \partial x_h^+}\delta x_j^+ \delta x_h^+ + \frac{\partial^2 V^+}{\partial x_j^+ \partial \lambda_h}\delta x_j^+ \delta \lambda_h + \frac{1}{2}\frac{\partial^2 V^+}{\partial \lambda_j \partial \lambda_h}\delta \lambda_j \delta \lambda_h\label{eq:DDP_quadratic_expansion_of_Vk1}
\end{align}
where
\begin{align}
\delta x_i^+ &:= x_i^+ - \bar{x}_i^+\\
&= f_i(\bm{x}, \bm{u}) - f_i(\bar{\bm{x}}, \bar{\bm{u}})
\end{align}
and $\delta x_i^+$ is obtained from Eq. (\ref{eq:DDP_quadratic_expansion_of_f}).

Therefore, substituting Eq. (\ref{eq:DDP_quadratic_expansion_of_Vk1}) to Eq. (\ref{eq:DDP_quadratic_expansion_of_f}) and adding Eq. (\ref{eq:DDP_quadratic_expansion_of_l}) yields the quadratic form of the right hand side of the Bellman equation (\ref{eq:DDP_bellman_equation}) as
\begin{equation}
\min_{\delta\bm{u}_k\in\delta\mathcal{U}_k} \left\{Q_0 + \begin{bmatrix}
\bm{Q}_x^T & \bm{Q}_u^T & \bm{Q}_{\lambda}^T
\end{bmatrix}\delta\bm{y}_k + \frac{1}{2} \delta\bm{y}_k^T\begin{bmatrix}
\textrm{Q}_{xx} & \textrm{Q}_{xu} & \textrm{Q}_{x\lambda}\\
\textrm{Q}_{xu}^T & \textrm{Q}_{uu} & \textrm{Q}_{u\lambda}\\
\textrm{Q}_{x\lambda}^T & \textrm{Q}_{u\lambda}^T & \textrm{Q}_{\lambda\lambda}
\end{bmatrix}\delta\bm{y}_k
\right\}\label{eq:DDP_quadratically_expanded_Bellman_equation}
\end{equation}
where we here remove the Cartesian tensor notation, $\delta\mathcal{U}_k$ is a set of admissible control variations, and the coefficients of the quadratic form are derived as follows
\begin{align}
Q_0 &= l(\bar{\bm{x}}_k, \bar{\bm{u}}_k) + V^+(\bar{\bm{x}}_{k+1}, \bar{\bm{\lambda}})\\
\bm{Q}_x &= \frac{\partial l}{\partial x_j} + \frac{\partial V^+}{\partial x_i^+}\frac{\partial f_i}{\partial x_j}\\
\bm{Q}_u &= \frac{\partial l}{\partial u_j} + \frac{\partial V^+}{\partial x_i^+}\frac{\partial f_i}{\partial u_j}\\
\bm{Q}_{\lambda} &= \frac{\partial V^+}{\partial \lambda_j}\\
\textrm{Q}_{xx} &= \frac{\partial^2 l}{\partial x_j \partial x_h} + \frac{\partial^2 V^+}{\partial x_m^+ \partial x_n^+}\frac{\partial f_m}{\partial x_j}\frac{\partial f_n}{\partial x_h} + \frac{\partial V^+}{\partial x_i^+}\frac{\partial^2 f_i}{\partial x_j\partial x_h}\\
\textrm{Q}_{xu} &= \frac{\partial^2 l}{\partial x_j \partial u_h} + \frac{\partial^2 V^+}{\partial x_m^+ \partial x_n^+}\frac{\partial f_m}{\partial x_j}\frac{\partial f_n}{\partial u_h} + \frac{\partial V^+}{\partial x_i^+}\frac{\partial^2 f_i}{\partial x_j\partial u_h}
\end{align}
\begin{align}
\textrm{Q}_{x\lambda} &= \frac{\partial^2 V^+}{\partial x_m^+ \partial \lambda_h}\frac{\partial f_m}{\partial x_j}\\
\textrm{Q}_{uu} &= \frac{\partial^2 l}{\partial u_j \partial u_h} + \frac{\partial^2 V^+}{\partial x_m^+ \partial x_n^+}\frac{\partial f_m}{\partial u_j}\frac{\partial f_n}{\partial u_h} + \frac{\partial V^+}{\partial x_i^+}\frac{\partial^2 f_i}{\partial u_j\partial u_h}\\
\textrm{Q}_{u\lambda} &= \frac{\partial^2 V^+}{\partial x_m^+ \partial \lambda_h}\frac{\partial f_m}{\partial u_j}\\
\textrm{Q}_{\lambda\lambda} &= \frac{\partial^2 V^+}{\partial \lambda_j \partial \lambda_h}
\end{align}

\subsection{Backward Sweep}
The backward sweep solves the quadratically expanded Bellman equation (\ref{eq:DDP_quadratically_expanded_Bellman_equation}). Calculating the derivative with respect to $\delta \bm{u}_k$ and nullifying it derive the optimal control correction as
\begin{equation}
\delta \bm{u}_k^* = \bm{\alpha}_k + \beta_k \delta \bm{x}_k + \gamma_k \delta \bm{\lambda}.\label{eq:DDP_optimal_control_correction}
\end{equation}
where
\begin{align}
\bm{\alpha}_k &= - \textrm{Q}_{uu}^{-1}\bm{Q}_u\\
\beta_k &= - \textrm{Q}_{uu}^{-1} \textrm{Q}_{xu}^T\\
\gamma_k &= - \textrm{Q}_{uu}^{-1} \textrm{Q}_{u\lambda}.
\end{align}
$\textrm{Q}_{uu}$ must be invertible; however, this criterion lacks universality within nonlinear optimal control problems. Among techniques to avoid the singularity\cite{Jacobson1970, Liao1991, Lantoine2012p1, Lantoine2012p2}, we adopt a trust-region method\cite{Conn2000, Lantoine2012p1, Lantoine2012p2}.

The backward sweep also compute the partial derivatives of $V_{k+1}^*(\cdot)$ recursively. Expanding the left hand side of the Bellman equation (\ref{eq:DDP_bellman_equation}) by the second order yields
\begin{align}
V_{k}^{*}(\bm{x}_k, \bm{\lambda}) &\simeq V_k^*(\bar{\bm{x}}_k, \bar{\bm{\lambda}}) + \left(\frac{\partial V_{k}^{*}}{\partial \bm{x}_k}\right)^T \delta\bm{x}_k + \left(\frac{\partial V_{k}^{*}}{\partial \bm{\lambda}}\right)^T \delta\bm{\lambda} \nonumber\\
& \ \ \ + \frac{1}{2} \delta\bm{x}_k^T \frac{\partial^2 V_{k}^{*}}{\partial \bm{x}_k^2} \delta\bm{x}_k + \delta\bm{x}_k^T \frac{\partial^2 V_{k}^{*}}{\partial \bm{x}_k \partial\bm{\lambda}} \delta\bm{\lambda} + \frac{1}{2} \delta\bm{\lambda}^T \frac{\partial^2 V_{k}^{*}}{\partial \bm{\lambda}^2} \delta\bm{\lambda}.\label{eq:DDP_quadratic_expansion_of_Vk}
\end{align}
where the partial derivatives are evaluated at $\bar{\bm{x}}_k$ and $\bar{\bm{\lambda}}$.

Substituting Eq. (\ref{eq:DDP_optimal_control_correction}) to Eq. (\ref{eq:DDP_quadratically_expanded_Bellman_equation}) and comparing the terms with Eq. (\ref{eq:DDP_quadratic_expansion_of_Vk}) provide the partial derivatives of $V_{k}^*(\cdot)$ as

\begin{align}
V_k^*(\bar{\bm{x}}_k, \bar{\bm{\lambda}}) &= Q_0 + \bm{Q}_u^T \bm{\alpha}_k + \frac{1}{2}\bm{\alpha}_k^T \textrm{Q}_{uu} \bm{\alpha}\\
\frac{\partial V_{k}^{*}}{\partial \bm{x}_k} &= \bm{Q}_x + \beta_k^T \bm{Q}_u + \textrm{Q}_{xu} \bm{\alpha}_k + \beta_k^T \textrm{Q}_{uu} \bm{\alpha}_k\\
\frac{\partial V_{k}^{*}}{\partial \bm{\lambda}} &= \bm{Q}_{\lambda} + \gamma_k^T \bm{Q}_u + \textrm{Q}_{u\lambda}^T \bm{\alpha}_k + \gamma_k^T \textrm{Q}_{uu} \bm{\alpha}_k\\
\frac{\partial^2 V_{k}^{*}}{\partial \bm{x}_k^2} &= \textrm{Q}_{xx} + \textrm{Q}_{xu}\beta_k + \beta_k^T \textrm{Q}_{xu}^T + \beta_k^T \textrm{Q}_{uu} \beta_k\\
\frac{\partial^2 V_{k}^{*}}{\partial \bm{x}_k \partial \bm{\lambda}} &= \textrm{Q}_{x\lambda} + \textrm{Q}_{u\lambda}^T\beta_k + \gamma_k^T \textrm{Q}_{xu}^T + \beta_k^T \textrm{Q}_{uu} \gamma_k\\
\frac{\partial^2 V_{k}^{*}}{ \partial \bm{\lambda}^2} &= \textrm{Q}_{\lambda\lambda} + \textrm{Q}_{u\lambda}^T \gamma_k + \gamma_k^T \textrm{Q}_{u\lambda} + \gamma_k^T \textrm{Q}_{uu} \gamma_k
\end{align}

The terminal conditions of $V_{N+1}^*(\bar{\bm{x}}_{N+1}, \bar{\bm{\lambda}})$ and its derivatives are given by the terminal cost $\tilde{\phi}(\bm{x}_{N+1}, \bm{\lambda})$ as
\begin{align}
V_{N+1}^*(\bar{\bm{x}}_{N+1}, \bar{\bm{\lambda}}) &= \tilde{\phi}(\bar{\bm{x}}_{N+1}, \bar{\bm{\lambda}})\\
\frac{\partial V_{N+1}^{*}}{\partial \bm{x}_{N+1}} &= \frac{\partial \tilde{\phi}}{\partial \bm{x}_{N+1}}\\
\frac{\partial V_{N+1}^{*}}{\partial \bm{\lambda}} &= \bm{\psi}(\bar{\bm{x}}_{N+1})\\
\frac{\partial^2 V_{N+1}^{*}}{\partial \bm{x}_{N+1}^2} &= \frac{\partial^2 \tilde{\phi}}{\partial \bm{x}_{N+1}^2}\\
\frac{\partial^2 V_{N+1}^{*}}{\partial \bm{x}_{N+1} \partial \bm{\lambda}} &= \frac{\partial \bm{\psi}}{\partial \bm{x}_{N+1}}\\
\frac{\partial^2 V_{N+1}^{*}}{\partial \bm{\lambda}^2} &= \bm{0}.
\end{align}
where the partial derivatives are evaluated at $\bar{\bm{x}}_{N+1}, \bar{\bm{\lambda}}$.

\subsection{Forward Sweep}

The forward sweep updates the reference trajectory $\{\bar{\bm{x}}_k\}_{k\in\mathbb{N}_{N+1}}$, $\{\bar{\bm{u}}_k\}_{k\in\mathbb{N}_{N}}$, and $\bar{\bm{\lambda}}$ by propagating trajectory with the control policy $\{\bm{\mu}_k(\bm{x}_k, \bm{\lambda})\}_{k\in\mathbb{N}_{N}}$ where
\begin{equation}
\bm{\mu}_k(\bm{x}_k) := \bar{\bm{u}}_k + \bm{\alpha}_k + \beta_k (\bm{x}_k-\bar{\bm{x}}_k) + \gamma_k (\bm{\lambda}-\bar{\bm{\lambda}}), \ \ \ k\in\mathbb{N}_{N}.
\end{equation}

Let us start with describing the initial stage of the propagation and explain the generic $k$-th stage. Given initial condition $\bar{\bm{x}}_0$ and the updated Lagrange multiplier $\bm{\lambda}$, the initial state and control vectors are updated as
\begin{align}
\bm{x}_0 &= \bar{\bm{x}}_0\\
\bm{u}_0 &= \bar{\bm{u}}_0 + \bm{\alpha}_0 + \gamma_0 (\bm{\lambda}-\bar{\bm{\lambda}}).
\end{align}
The dynamical system (\ref{eq:DDP_dynsys}) propagates the trajectory and the state vector $\bm{x}_1$ is obtained as
\begin{equation}
\bm{x}_1 = \bm{f}_0(\bm{x}_0, \bm{u}_0).
\end{equation}
Here, $\bm{x}_1$ is the updated state vector at 1-st stage.

For generic $k$-th stage, $\bm{x}_k$ is given by applying the aforementioned process recursively, and the $k$-th stage control vector is updated as
\begin{align}
\bm{u}_k &= \bar{\bm{u}}_k + \bm{\alpha}_k + \beta_k (\bm{x}_k-\bar{\bm{x}}_k) + \gamma_k (\bm{\lambda}-\bar{\bm{\lambda}})
\end{align}
The dynamical system (\ref{eq:DDP_dynsys}) propagates the trajectory and the state vector $\bm{x}_{k+1}$ is obtained as
\begin{equation}
\bm{x}_{k+1} = \bm{f}_k(\bm{x}_k, \bm{u}_k).
\end{equation}

The forward sweep process should also evaluate the cost reduction. The expected cost reduction at $k$-th stage can be computed by
\begin{align}
ER_k &:= V_k^*(\bar{\bm{x}}_k) - Q_0\\
&= \bm{Q}_u^T \bm{\alpha}_k + \frac{1}{2}\bm{\alpha}_k^T \textrm{Q}_{uu} \bm{\alpha}
\end{align}
where our implementation computes $ER_k$ in the backward sweep. The total expected cost reduction $ER$ is therefore
\begin{equation}
ER = \sum_{k=0}^{N} ER_k.
\end{equation}
We compare this expected cost reduction with the actual cost reduction obtained through the updated trajectory, and evaluate the convergence of the optimization problem.

%
%

\subsection{Sundman Transformation}

Aziz et al.\cite{Aziz2017} have shown that incorporating a Sundman transformation, which shifts the independent variable from the time domain to an orbital angle domain, results in a notable improvement in the effectiveness of the HDDP methodology for designing low-thrust, many-revolution trajectories. Since the transformation from the time domain to the true anomaly domain shows the most promising results according to their results, this paper adopts the Sundman transformation from $t$ to the true anomaly $\nu$
\begin{equation}
    dt = \frac{r^2}{h} d\nu
\end{equation}
where $r$ is the orbital radius and $h$ is the norm of the angular momentum, i.e., $h:=\|\bm{r}\times\bm{v}\|$.

Thus, the time derivative of a variable $z$ is replaced by
\begin{equation}
    \frac{dz}{dt} = \frac{dz}{d\nu}\frac{d\nu}{dt} = \frac{dz}{d\nu}\frac{h}{r^2}
\end{equation}

\section{Spiral Orbit-Raising Model}

\subsection{Dynamical System and Stage Constraints}

This paper deals with a point-mass gravity model of the Earth without considering the gravity of the Moon or the oblateness of the Earth. For the state vector $\boldsymbol{x} = \left[\boldsymbol{r}^{\top} \ \boldsymbol{v}^{\top} \ m \right]^{\top}$, the discrete-time dynamical system is calculated through the numerical integration of the equation of motion from $\nu$ to $\nu+\Delta \nu$:
\begin{equation}
\frac{d}{d\nu} \begin{bmatrix}\boldsymbol{r}\\ \boldsymbol{v}\\ m \end{bmatrix} = \frac{r^2}{h}\begin{bmatrix}
\boldsymbol{v}\\
-\frac{\mu}{\|\boldsymbol{r}\|^3} \boldsymbol{r} + \frac{\boldsymbol{u}}{m}\\
- \frac{\sqrt{\|\boldsymbol{u}\|^2-\epsilon_{leak}}}{g_0 I_{sp}}
\end{bmatrix}
\end{equation}
where $\epsilon_{leak}$ is a mass leak parameter ($\epsilon_{leak}=10^{-6}$ in our implementation), and the control inputs $\bm{u}_k$ are constrained by the upper bound of the thrust magnitude.
\begin{equation}
g_k(\boldsymbol{x}_k, \boldsymbol{u}_k) = \|\boldsymbol{u}_k\|^2 - u_{\textrm{UB}}^2 = 0
\end{equation}

\subsection{Phase Constraints}

To perform a lunar flyby, the terminal boundary condition specifies that the spacecraft's position with respect to the Moon is zero. If we ignore the phase of the moon, using the Keplerian orbital elements with respect to the lunar orbital plane, the lunar flyby occurs when the following condition is satisfied 
\begin{equation}
r_{\textrm{moon}} = \begin{cases}
\frac{p}{1+e\cos(-\omega)} \textrm{ \ \ \ if the ascending node crosses the moon}\\
\frac{p}{1+e\cos(\pi-\omega)} \textrm{ \ \ \ if the descending node crosses the moon}
\end{cases}
\end{equation}
where $p(=h^2/\mu)$ is the semi-latus rectum, $e$ is the eccentricity, $\omega$ is the argument of periapsis, and $r_{\textrm{moon}}$ is the orbital radius of the Moon. Of the two conditions, the choice of the node on the apogee side is preferable, so let us derive the general form of the termination boundary condition for the apogee side.

From the definition of the ascending node, the unit vector directed to the ascending node is
\begin{equation}
    \boldsymbol{n} = \frac{\bm{k}\times\bm{h}}{\|\bm{k}\times\bm{h}\|}
\end{equation}
where $\bm{k}$ is the unit vector in the direction of $z$ axis and $\bm{h}(=\bm{r}\times\bm{v})$ is the angular momentum vector. Given the vector $\bm{n}$, the cosine of the argument of periapsis is calculated
\begin{equation}
\cos\omega = \frac{\boldsymbol{e}\cdot\boldsymbol{n}}{e}
\end{equation}
where $\bm{e}$ is the eccentricity vector defined as
\begin{align}
\boldsymbol{e} := \frac{\boldsymbol{v}\times\boldsymbol{h}}{\mu} - \frac{\boldsymbol{r}}{r}.
\end{align}

The ascending node is on the apogee side when $\bm{e}\cdot\bm{n} < 0$, and the descending node is on the apogee side when $\bm{e}\cdot\bm{n} > 0$. Therefore, the general form of the crossing condition is written as
\begin{equation}
r_{\textrm{moon}} = \frac{p}{1 - |\boldsymbol{e}\cdot\boldsymbol{n}|}
\end{equation}
That is,
\begin{equation}
    \psi(\bm{x}_{N+1}) = \mu r_{\textrm{moon}}\left(1 - |\boldsymbol{e}\cdot\boldsymbol{n}|\right) - h^2 = 0.
\end{equation}

\subsection{Stage and Phase Cost}

In this low-thrust, many-revolution trajectory design problem, we maximize the final mass of the spacecraft, that is,
\begin{equation}
\phi(\bm{x}_{N+1}) = - m_{N+1}.
\end{equation}

After escaping the radiation belt, the spacecraft should not pass through it, so we define a penalty function for the perigee altitude. The penalty function is modeled by an exponential barrier function defiend as 
\begin{equation}
l_k(\boldsymbol{x}_k, \boldsymbol{u}_k) =\epsilon \exp\left(-\frac{\|\bm{r}_k\|-r_{\textrm{min}}}{\epsilon}\right)
\end{equation}
where $\epsilon$ is a small positive number ($\epsilon = 10^{-4}$ in our implementation).

\section{Numerical Simulation}

This section first shows the nominal low-thrust, many-revolution trajectory design without implementing uncertainties. To evaluate the effects of uncertainties, we perform the Monte Carlo simulation by randomly adding the initial state errors and the thrust execution errors. In this analysis, HDDP with Sundman transformation is implemented in Julia language using the differential equation module \texttt{DifferentialEquations.jl}\cite{JuliaDE} and the quadratic programming module \texttt{OSQP.jl}\cite{JuliaOSQP}.

\subsection{Nominal Trajectory Design}

This numerical simulation deals with the trajectory optimization during the spiral orbit-raising phase of \destiny\cite{Ozaki2022}. In the first half of the phase, \destiny continuously accelerates along the tangential direction of the orbit so that the spacecraft can escape from the radiation belt (altitude $<$ 20,000 km) as quickly as possible. In the latter half of the phase, we optimize a low-thrust, many-revolution trajectory considering arbitrary low-thrust directions and coasting arcs. The initial conditions of the following analyses are fixed to the state vector (i.e., position, velocity, and mass) immediately after the spacecraft exits the radiation belt. The simulation conditions are shown in Table \ref{tab:parameter_settings}.

\begin{table}[htb]
\centering
\caption{Parameter settings.}\label{tab:parameter_settings}
\begin{tabular}{llll}\hline\hline
Parameters & Variables & Settings & Units \\ \hline
Number of segments & $N$ & 6000$\sim$8000 & -\\
True anomaly step size & $\Delta \nu$ & $0.02\pi$ & rad \\
Maximum thrust magnitude & $u_{UB}$ & 40 & mN\\
Specific impulse & $I_{sp}$ & 3000 & s\\
Minimum orbital radius & $r_{\textrm{min}}$ & 26378.1366 & km\\
Moon's orbital radius & $r_{\textrm{moon}}$ &  384748.0 & km\\
Initial epoch & $t_0$ & 2025 MAR 02 13:46:16.920 & TDB\\
Initial position & $\bm{r}_0$ & $\begin{bmatrix}20360.65082405\\ 21215.73853905543\\ -30668.77526763988\end{bmatrix}$ & km \\
Initial velocity & $\bm{v}_0$ & $\begin{bmatrix}-1.92766723\\ 1.647683013442788\\ -2.253212251694917\end{bmatrix}$ & km/s\\
Initial mass & $m_0$ & 455.14851 & kg\\
Scale factors of position & $L_{\textrm{sf}}$ & $10^5$ & km\\
Scale factors of time & $T_{\textrm{sf}}$ & $10^5$ & s\\
Scale factors of mass & $M_{\textrm{sf}}$ & $10^2$ & kg\\\hline\hline
\end{tabular}
\end{table}%

\begin{figure}[h!]
\centering
\includegraphics[width=0.7\textwidth]{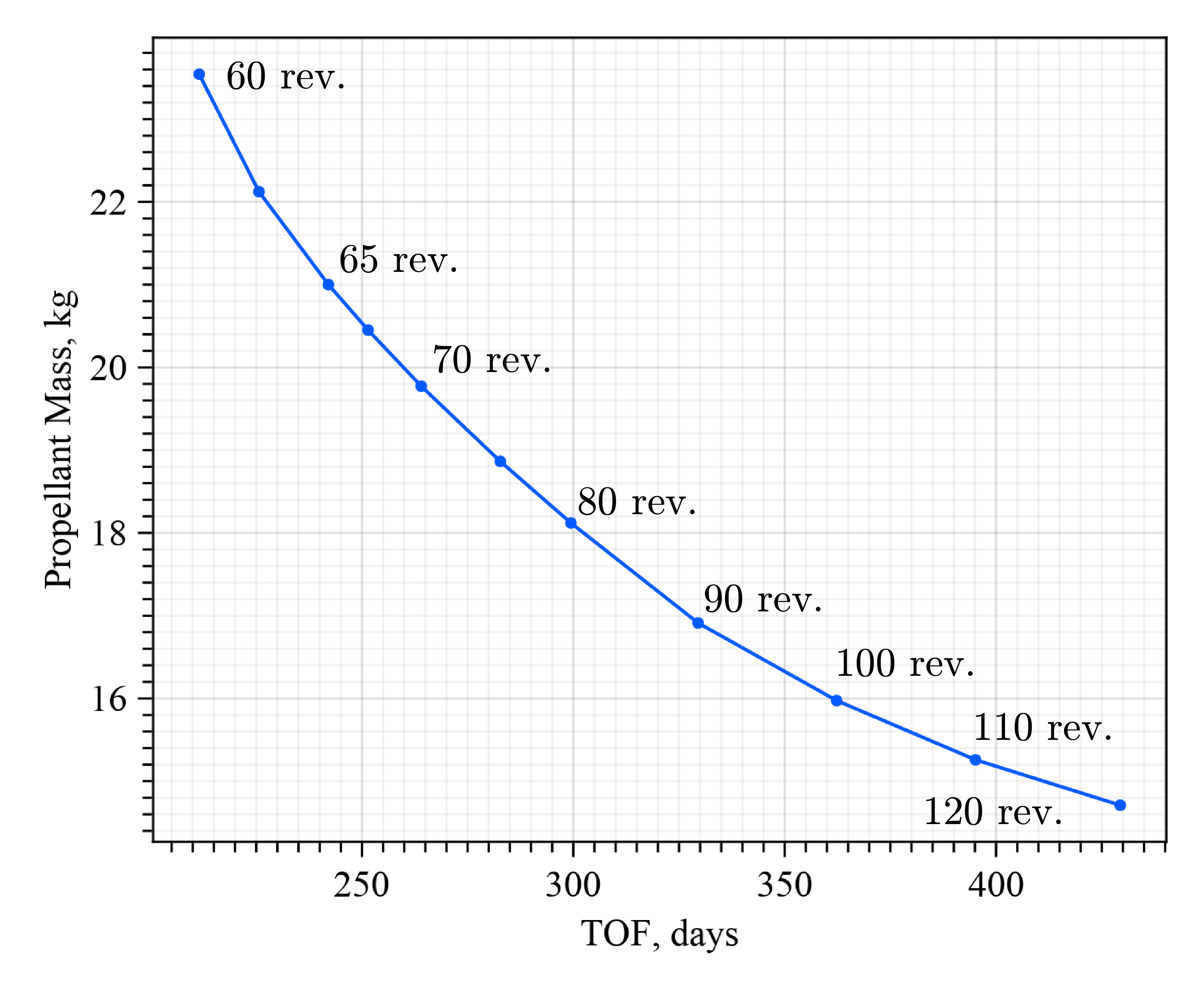}
\caption{Time of Flight (TOF) versus propellant consumption after leaving the radiation belt.}
\label{fig:tof_vs_mass}
\end{figure}

\begin{figure}[htbp]
	\begin{center}
		\begin{tabular}{cc}
			\hspace*{-0.075\hsize}
			\begin{minipage}{0.55\hsize}
				\begin{center}
					\includegraphics[clip,width=\hsize]{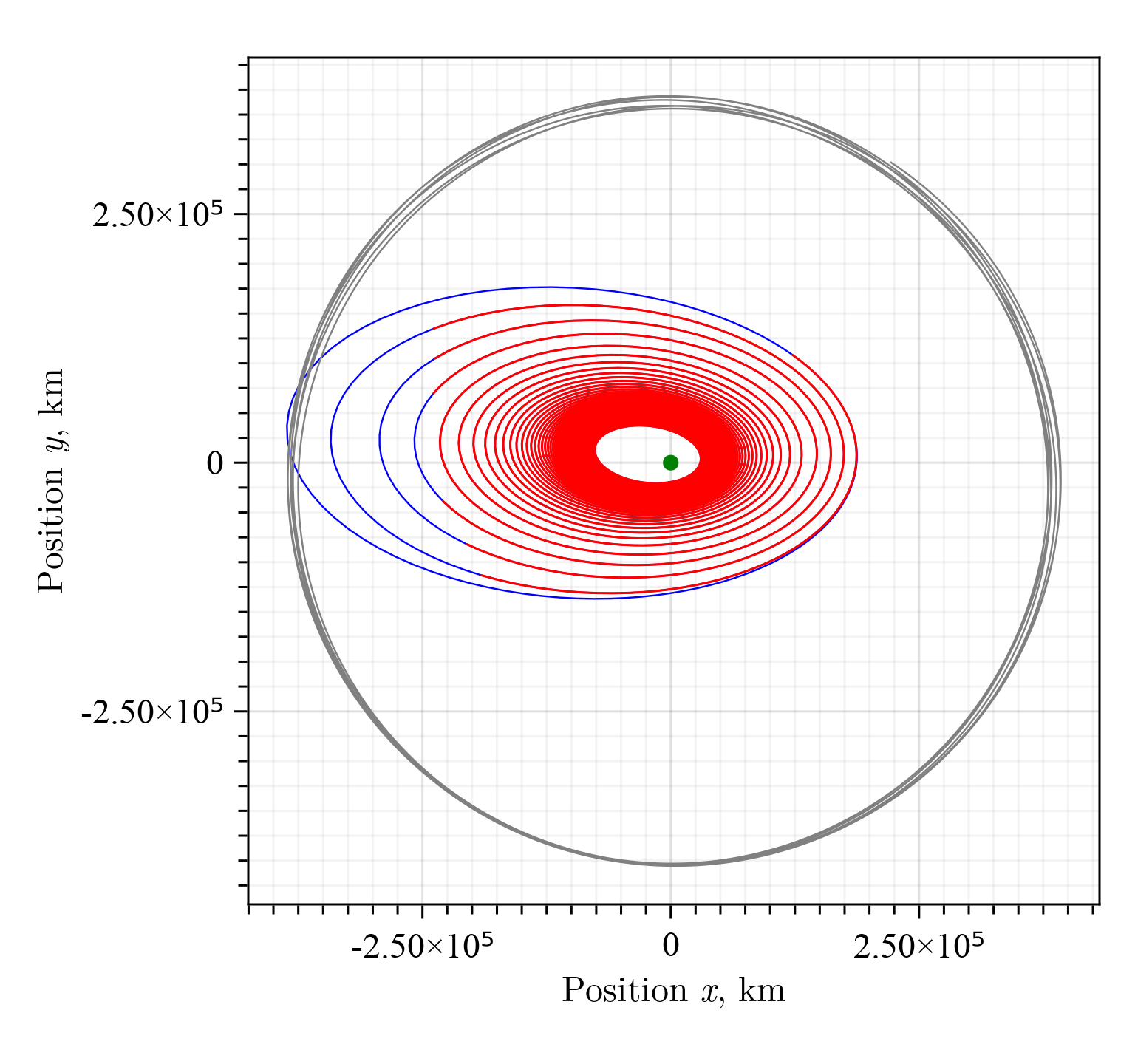}
					{\small a) Time-saving solution (60 rev., $N=6000$)}
					\vspace{7pt}
				\end{center}
			\end{minipage}
			\begin{minipage}{0.55\hsize}
				\begin{center}
					\includegraphics[clip,width=\hsize]{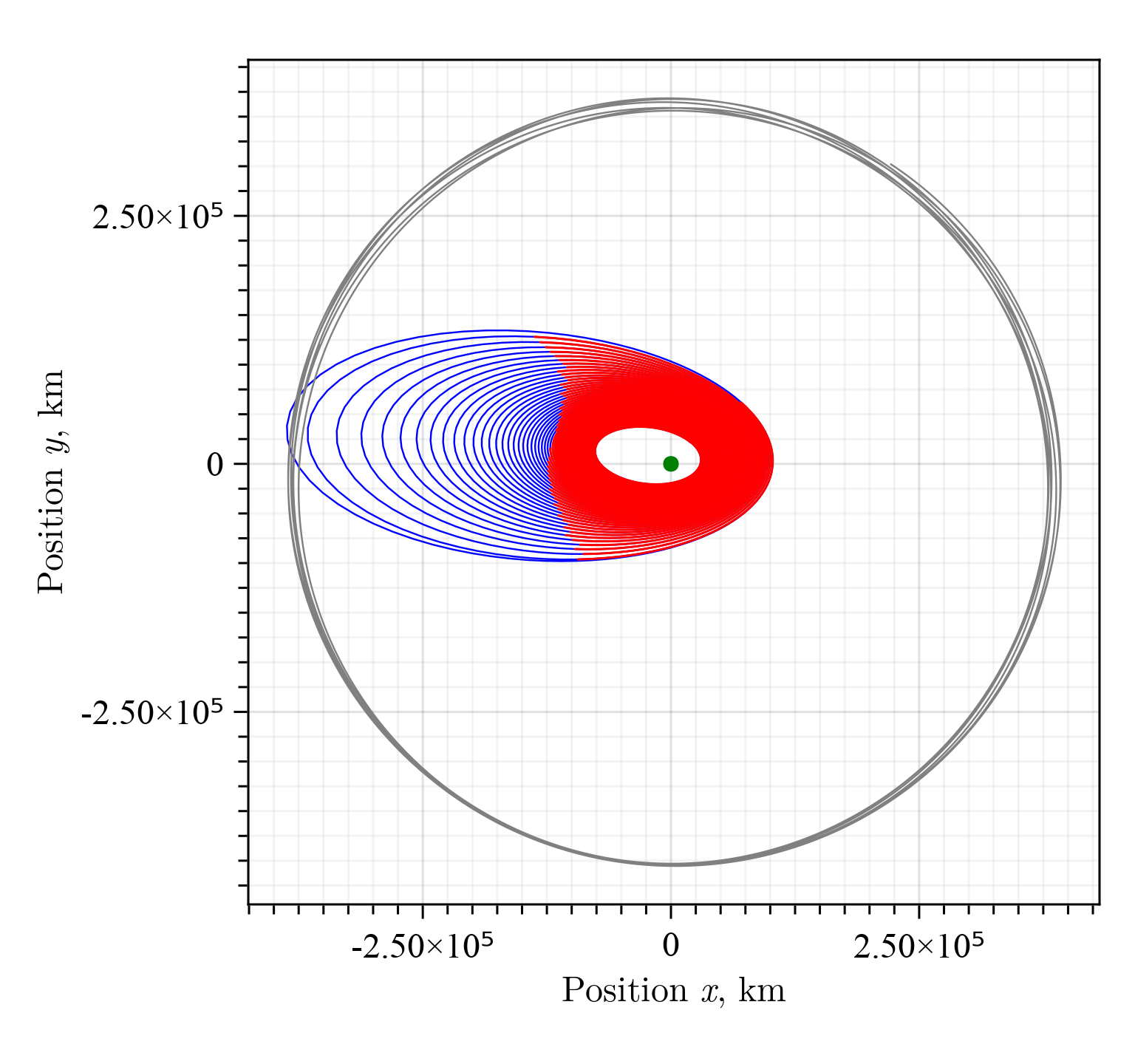}
					{\small b) Baseline solution (67 rev., $N=6700$)}
					\vspace{7pt}
				\end{center}
			\end{minipage}
		\end{tabular}
		\begin{tabular}{cc}
			\hspace*{-0.075\hsize}
			\begin{minipage}{0.55\hsize}
				\begin{center}
					\includegraphics[clip,width=\hsize]{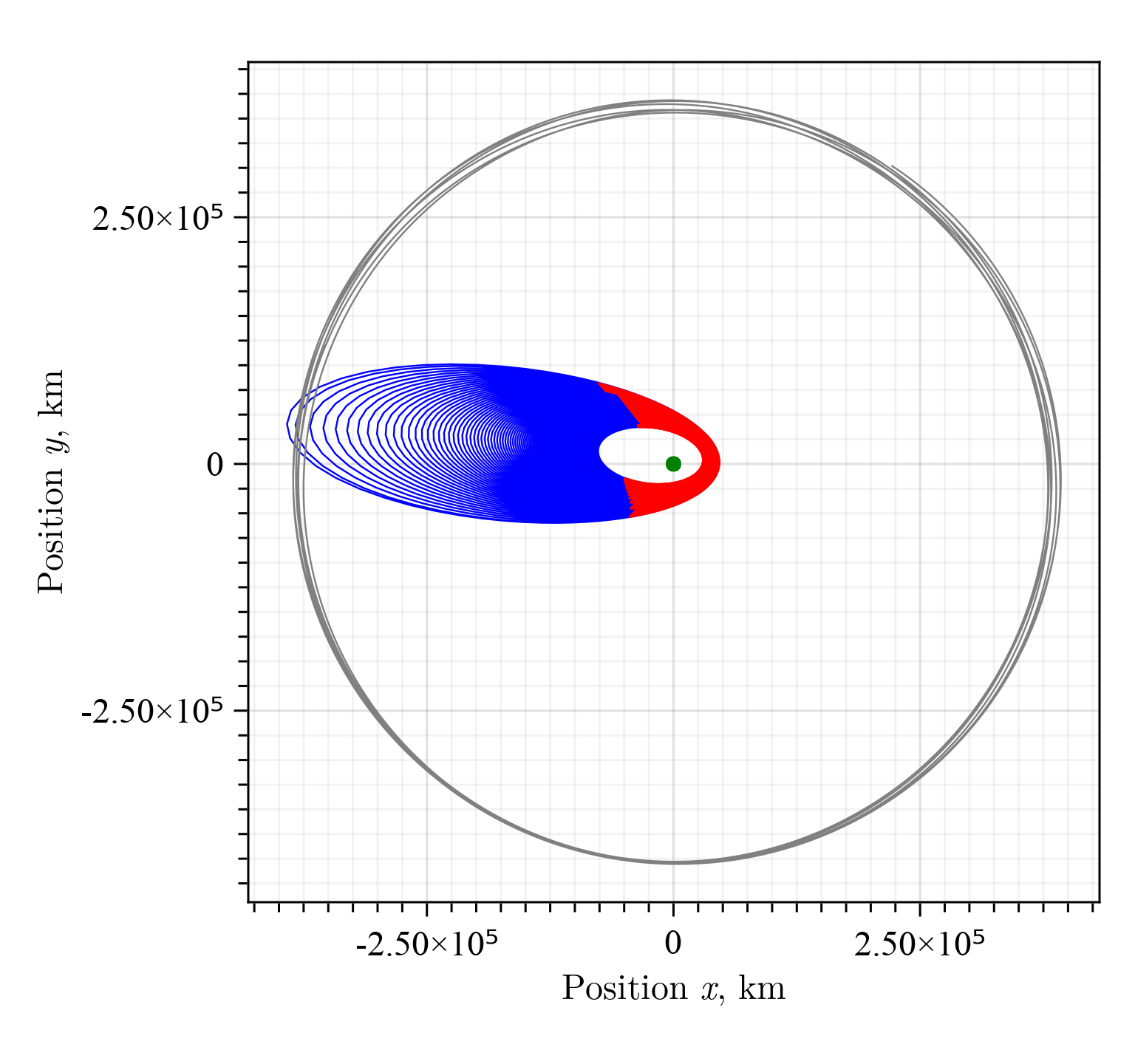}
					{\small c) Fuel-saving solution (120 rev., $N=12000$)}
					\vspace{7pt}
				\end{center}
			\end{minipage}
		\end{tabular}
		\caption{Nominal optimized trajectories (blue line represents the nominal trajectory, and red line shows the nominal thrusting arc).}
		\label{fig:nom_trj}
	\end{center}
\end{figure}

\destiny will orbit the Earth about 100 revolutions after leaving the radiation belt to reach the Moon, and the first lunar flyby must be performed on August 14, 2026, due to the Earth's escape conditions for the Phaethon flyby transfer. Therefore, there is a hard constraint on the flight time, which must be less than 530 days after leaving the radiation belt to reach the Moon. The spiral trajectory must be designed to allow enough time to get to the Moon, taking into account various uncertainties such as missed thrust and orbit determination errors. The propellant mass of the spacecraft is 60 kg for a wet mass of 480 kg, and the available propellant mass in the second half of the spiral phase is 23 kg. The spiral orbit-raising trajectory design problem is a multi-objective optimization problem, and the trade-off between fuel consumption and flight time is shown in Fig.\ref{fig:tof_vs_mass}. Three representative trajectories are shown in Fig.\ref{fig:nom_trj}.

\subsection{Monte Carlo Simulation}

We run the Monte Carlo simulations of orbit propagation with the uncertainties listed in Table \ref{tab:uncertainty_settings}. For each trajectory, we add an initial position error and an initial velocity error as a multivariate Gaussian random variable on the initial value. For each discretized step in the propagation, we also add the thrust errors as a Gaussian random variable. If the effect of the uncertainties is left uncorrected, the trajectory will gradually deviate and eventually the mission will fail. Therefore, to correct the uncertainties, we introduce the four different control policies:
\begin{enumerate}
\item Open-loop control in the time domain (=no uncertainty correction): $\bm{u}_k(t_k) = \bar{\bm{u}}_k$
\item Closed-loop control with DDP $\beta_k$ in the time domain: $\bm{u}_k(t_k) = \bar{\bm{u}}_k + \beta_k (\bm{x}_k(t_k)-\bar{\bm{x}}_k)$
\item Open-loop control in the angle domain: $\bm{u}_k(\nu_k) = \bar{\bm{u}}_k$
\item Closed-loop control with DDP $\beta_k$ in the angle domain: $\bm{u}_k(\nu_k) = \bar{\bm{u}}_k + \beta_k (\bm{x}_k(\nu_k)-\bar{\bm{x}}_k)$.
\end{enumerate}
where we modify the control input as $\bm{u}'_k = \left(u_{UB}/\|\bm{u}_k\|\right)\bm{u}_k$ if $\|\bm{u}_k\| > u_{UB}$. In cases 1 and 2, the spacecraft trajectory is propagated in the time domain; in cases 3 and 4, the spacecraft trajectory is propagated in the angle domain by the Sundman transform.

\begin{table}[htb]
\centering
\caption{Error settings.}\label{tab:uncertainty_settings}
\begin{tabular}{lll}\hline\hline
Parameters & Settings & Units \\ \hline
Initial position errors & 1 & km (1$\sigma$)\\
Initial velocity errors & 0.1 & m/s (1$\sigma$)\\
Instantaneous thrust errors & 0.7 & mN (1$\sigma$)\\\hline\hline
\end{tabular}
\end{table}%

\begin{figure}[htbp]
	\begin{center}
		\begin{tabular}{cc}
			\hspace*{-0.075\hsize}
			\begin{minipage}{0.55\hsize}
				\begin{center}
					\includegraphics[clip,width=\hsize]{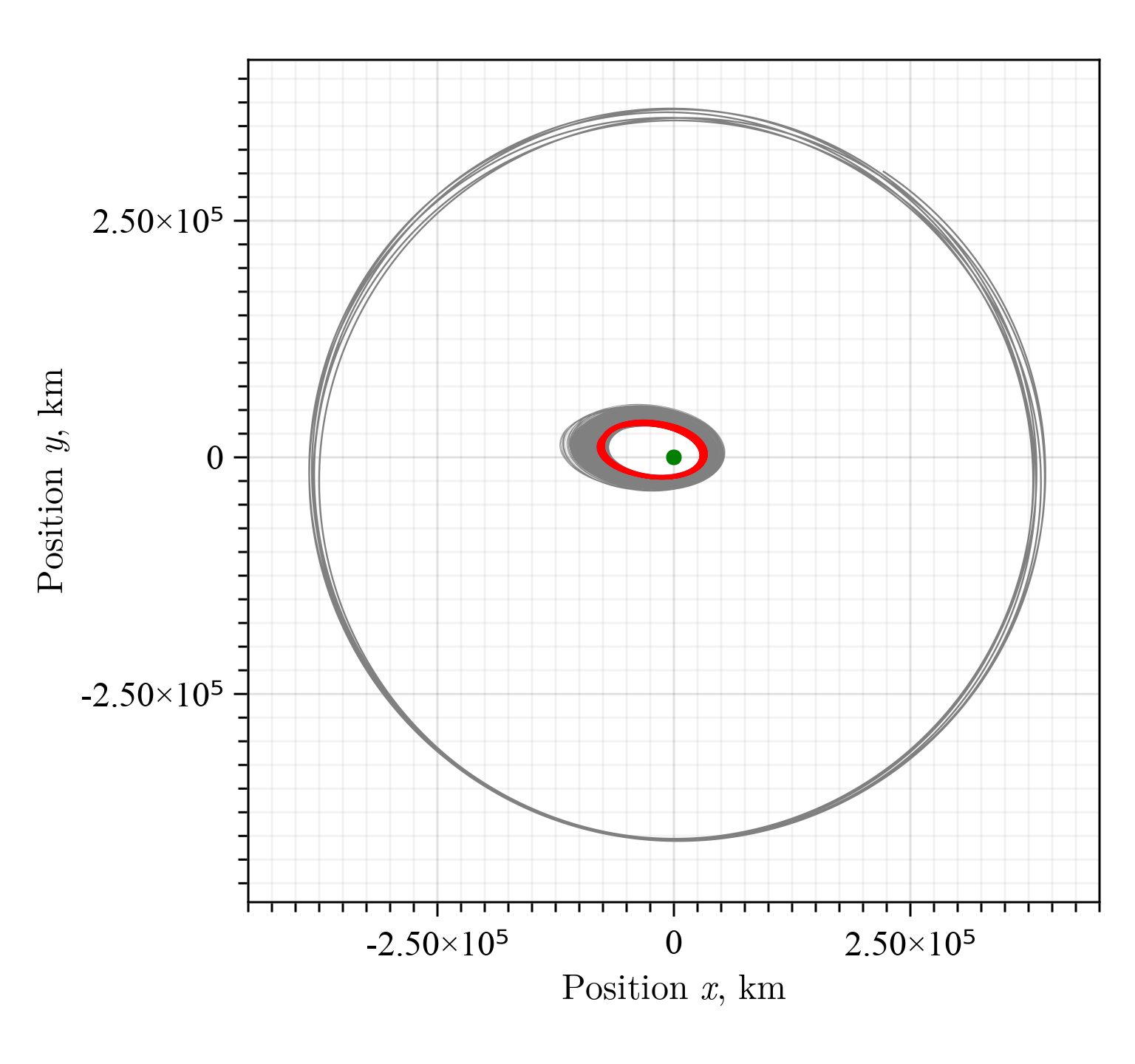}
					{\small a) Case 1: Open-loop control in time domain}
					\vspace{7pt}
				\end{center}
			\end{minipage}
			\begin{minipage}{0.55\hsize}
				\begin{center}
					\includegraphics[clip,width=\hsize]{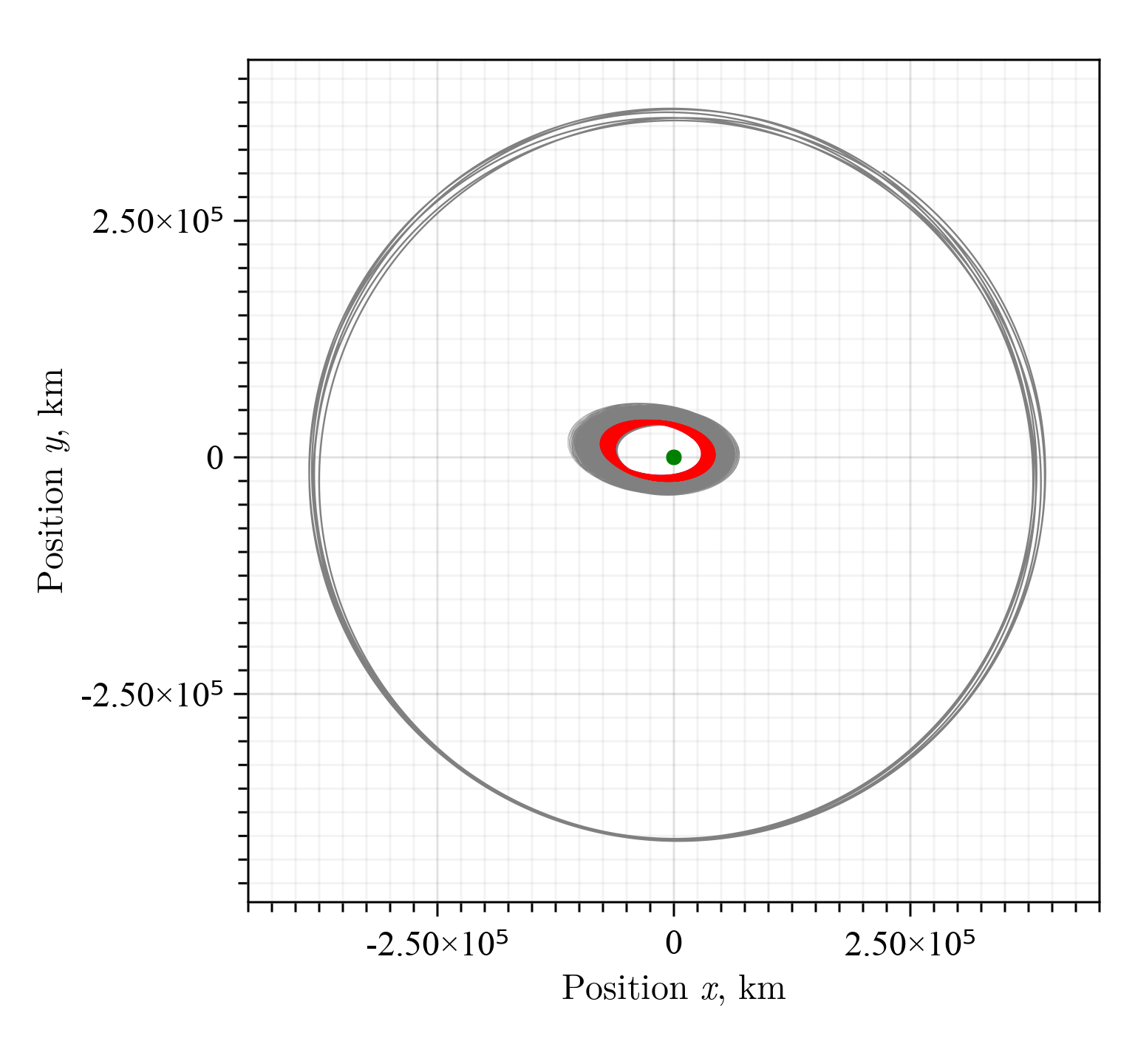}
					{\small b) Case 2: Closed-loop control of DDP in time domain}
					\vspace{7pt}
				\end{center}
			\end{minipage}
			\\
			\hspace*{-0.075\hsize}
			\begin{minipage}{0.55\hsize}
				\begin{center}
					\includegraphics[clip,width=\hsize]{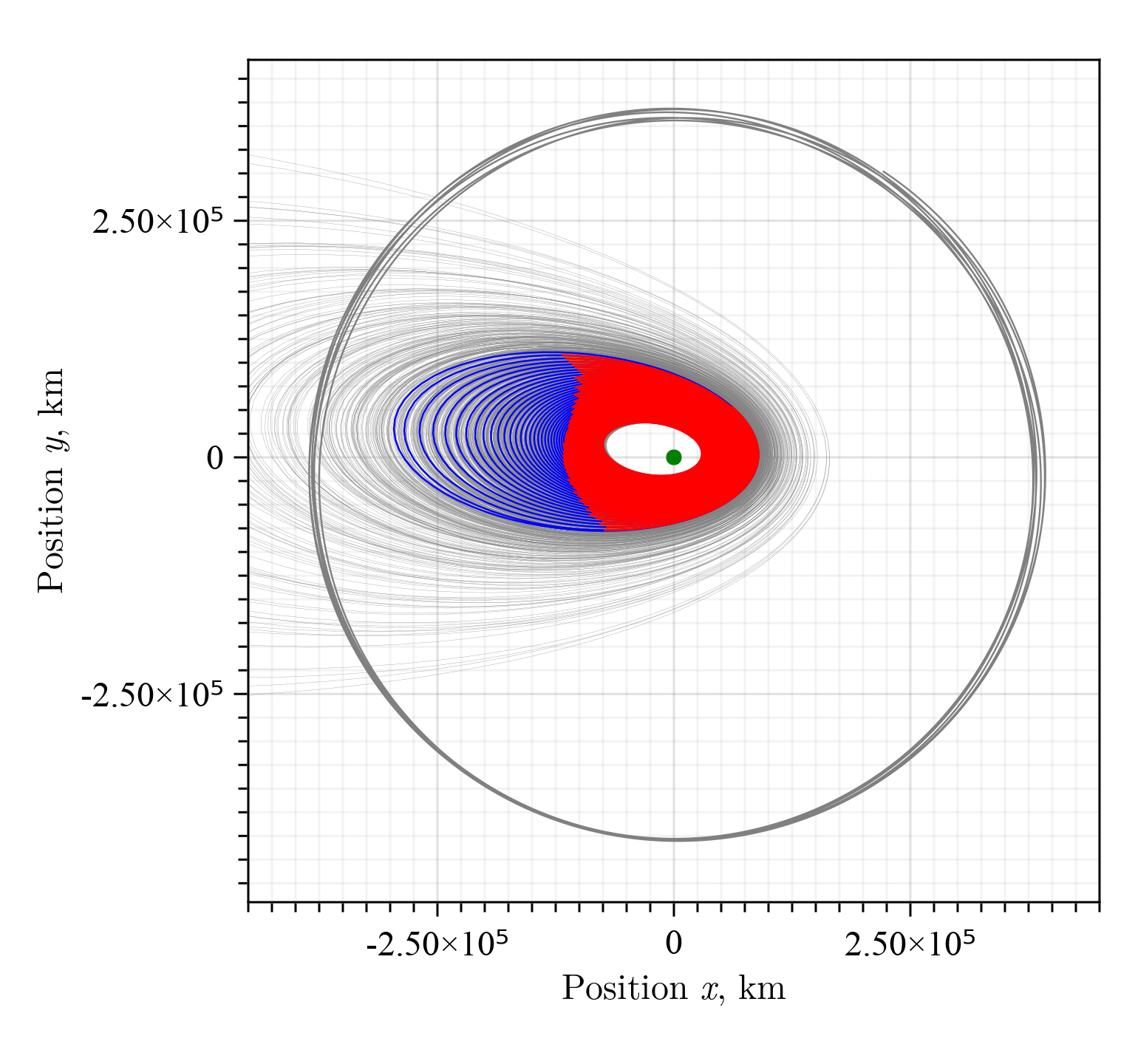}
					{\small c) Case 3:  Open-loop control in angle domain}
				\end{center}
			\end{minipage}
			\begin{minipage}{0.55\hsize}
				\begin{center}
					\includegraphics[clip,width=\hsize]{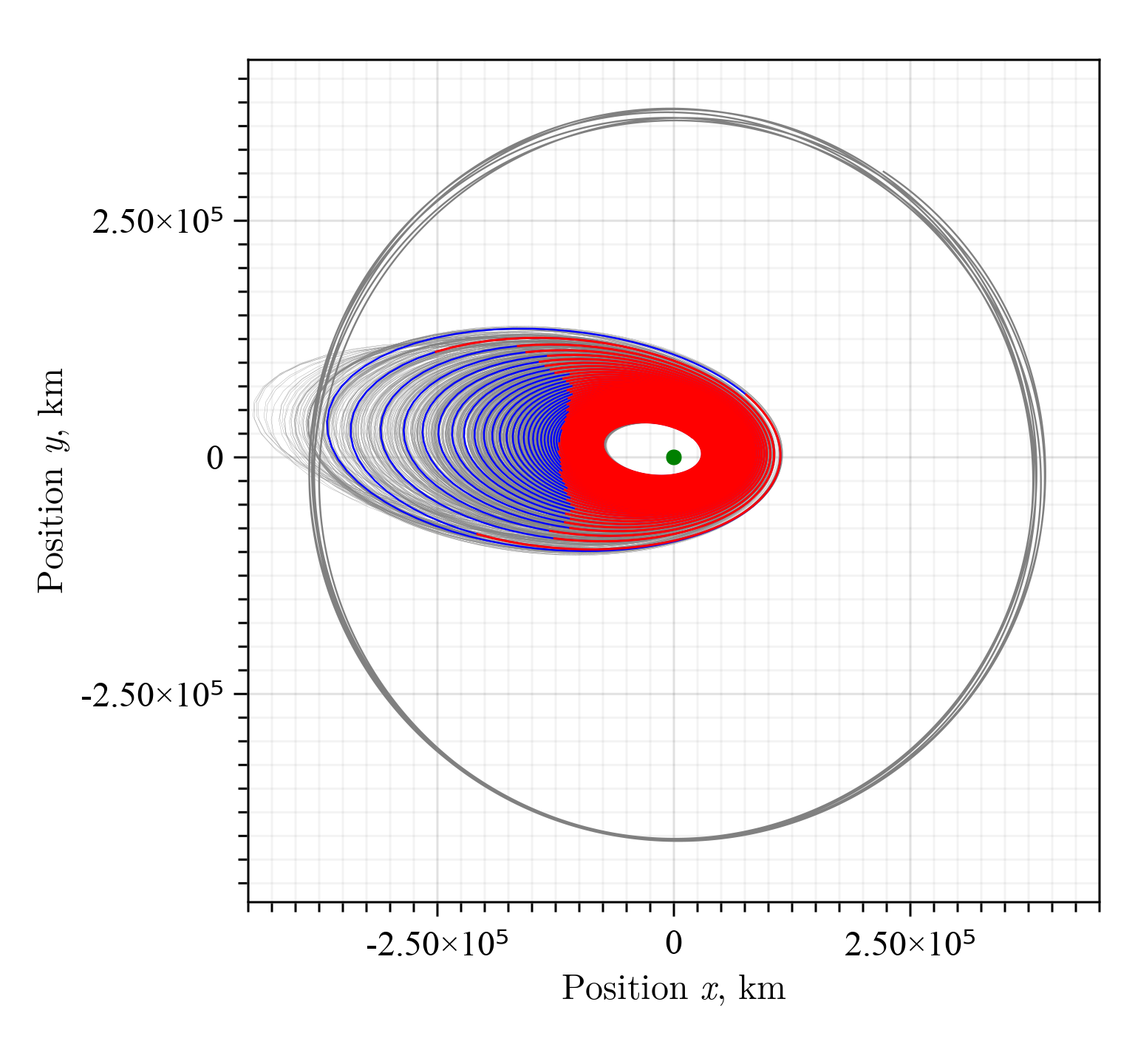}
					{\small d) Case 4: Closed-loop control of DDP in angle domain}
				\end{center}
			\end{minipage}
		\end{tabular}
		\caption{Perturbed trajectories in Monte Carlo simulation (50 trajectories, where each gray line indicates a sample trajectory, blue line represents a representative trajectory, and red line shows the representative thrusting arc).}
		\label{fig:mc_all}
	\end{center}
\end{figure}

The Monte Carlo simulation results in the perturbed trajectory shown in Fig.\ref{fig:mc_all}. The control policies in the angular domain, as shown in cases 3 and 4, dramatically improve the performance of the trajectory correction against uncertainties. In particular, in case 4, the perturbed trajectory reaches the moon as well as the nominal trajectory, as shown in Fig.\ref{fig:nom_trj} d). The resulting control policy allows the spacecraft to be guided to the lunar vicinity without iteratively optimizing the low-thrust, many-revolution trajectory.

\section{Discussion}

The Monte Carlo simulation results confirmed that the feedback control of the DDP solved with the Sundman transform can be practically used as the control law without having to re-optimize the low-thrust trajectory for a long period of time. Low-thrust many-revolution trajectories are not only difficult to design, but also pose great challenges to flight operations. The use of DDP feedback control in the angular domain will greatly contribute to the autonomy of low-thrust flight operations. Although the current uncertainty model does not introduce missed thrust, missed thrust can be easily evaluated as a time delay. If the mission designers allocate a sufficient time margin for the lunar flyby, the spacecraft should be able to reach the Moon robustly. In addition, since this study does not consider the phase of the lunar flyby, some further studies are needed to robustly control the trajectories for realistic missions. We have confirmed that feedback control in the angular domain is effective, and therefore the construction of surrogate-based feedback control laws in the angular domain using machine learning\cite{Izzo2021}, such as Guided Policy Search\cite{Levine2013}, is more likely to be generalizable.



\section{Conclusions}

Low-thrust, many-revolution trajectory design, considered one of the most challenging problems, expands the possibilities of space missions. In this paper, we take the spiral-orbit-raising phase of JAXA's \destiny mission as an example and design the low-thrust trajectory from Earth orbit to lunar transfer orbit by the Sundman-transformed HDDP. This paper performs Monte Carlo runs to simulate the operational uncertainties along the spiral orbit and confirms that the feedback control law of DDP in the angular domain can successfully guide most of the spacecraft trajectories to the Moon. These results contribute not only to the evaluation of the robustness of the trajectory but also to the autonomy of the flight operation.

\bibliographystyle{AAS_publication}   
\bibliography{references}   

\end{document}